# Engineering superconducting contacts transparent to a bipolar graphene

Seong Jang[1], Geon-Hyoung Park[1], Sein Park[1], Hyeon-Woo Jeong[1], Kenji Watanabe[2], Takashi Taniguchi[3] and Gil-Ho Lee[1]

[1] Department of Physics, Pohang University of Science and Technology, Pohang, Republic of Korea

[2] Research Center for Functional Materials, National Institute for Materials Science, Tsukuba, Japan

[3] International Center for Materials Nanoarchitectonics, National Institute for Materials Science, Tsukuba, Japan

Graphene's exceptional electronic mobility, gate-tunability, and contact transparency with superconducting materials make it ideal for exploring the superconducting proximity effect. However, the work function difference between graphene and superconductors causes unavoidable doping of graphene near contacts, forming a p-n junction in the hole-doped regime and reducing contact transparency. This challenges the device implementation that exploits graphene's bipolarity. To address this limitation, we developed a new fabrication scheme for two-dimensional superconducting contacts that allows independent control over charge concentration and polarity for both the graphene in contact with superconductors and the graphene channel. Contact transparency, conductance enhancement, and Josephson coupling were measured to confirm transparent contacts to both polarities of graphene. Moreover, we demonstrated the Andreev process in the quantum Hall edge state at a negative filling factor of $\nu = -2$. This scheme will open avenues for realizing various theoretical propositions utilizing the bipolarity of graphene combined with superconductivity.

A transparent superconducting contact is essential for achieving various exotic phenomena arising from the superconducting proximity effect. Previous theoretical and experimental works have demonstrated that a more transparent superconductor/normal (S/N) interface allows superconducting correlations to transfer more effectively from the superconductor to the normal conductor[1]. Graphene has emerged as a suitable platform for investigating the superconducting proximity effect due to its high mobility, gate-tunability, and zero-bandgap, which helps to avoid the Schottky barrier at metal contacts. Numerous noteworthy experimental works have demonstrated superconducting correlations within graphene, manifested in various forms such as Josephson junctions[2,3], specular Andreev reflection[4], crossed Andreev reflection[5] and superconducting correlation[6,7,8,9] and supercurrent[10,11] in the quantum Hall regime. These achievements are attributed to the development of the dry transfer method and one-dimensional contact[12](Figure 1a), ensuring high mobility of graphene and highly transparent contacts, respectively. However, the majority of the studies on superconducting proximity effect on graphene explore only negatively doped (n-doped) region due to low transparency in the positively doped (p-doped) region. Most of known superconducting material with high transparency to graphene – titanium (Ti) [6,13,14], tantalum (Ta)[2], niobium titanium nitride (NbTiN)[15] and molybdenum/rhenium alloy (MoRe)[10] – induces strong negative doping in graphene near the contact due to the work function difference between superconducting material and graphene. As depicted in Figure 1b, the Fermi level ($E_F$) near the edge contact is fixed positive regardless of that in the graphene channel. Thus, a p-n junction forms near the contact when the graphene is positively doped (p-doped), resulting in low contact transparency and weak superconducting proximity effect.

This barrier near the superconducting contact affects the efficiency of Andreev reflection. For example, the conductance enhancement in n-doped graphene exceeds 40%[6], while that in p-doped graphene is much lower and even shows a conductance dip near zero bias voltage (Figure S1). This issue becomes more severe in the quantum Hall regime as highly insulating phase of filling factor $\nu = 0$ forms at p-n junction, electrically disconnecting graphene from superconducting contacts. The inferior contact quality also hinders the exploration of various device schemes that leverage the bipolarity of graphene. For instance, a theoretical proposition[16] suggests connecting two oppositely spin-polarized states, $\nu = 1$ and $\nu = -1$, through a superconducting electrode to achieve perfect crossed Andreev reflection.

To overcome this problem, here we propose a dual-gated two-dimensional (2D) contact scheme[17,18](Figure 1c), where the charge concentration and the polarity of graphene layer in contact with superconductors are gate-tunable (Figure 1d). By minimizing the potential difference between the graphene channel and graphene in contact with the superconductor, we achieved high Andreev reflection probability as well as large Josephson coupling in both n-doped and p-doped graphene. Furthermore, we observed a pronounced negative downstream resistance in the quantum Hall state of negative filling factor $\nu = -2$ for the first time, exhibiting crossed Andreev conversion in p-doped graphene.

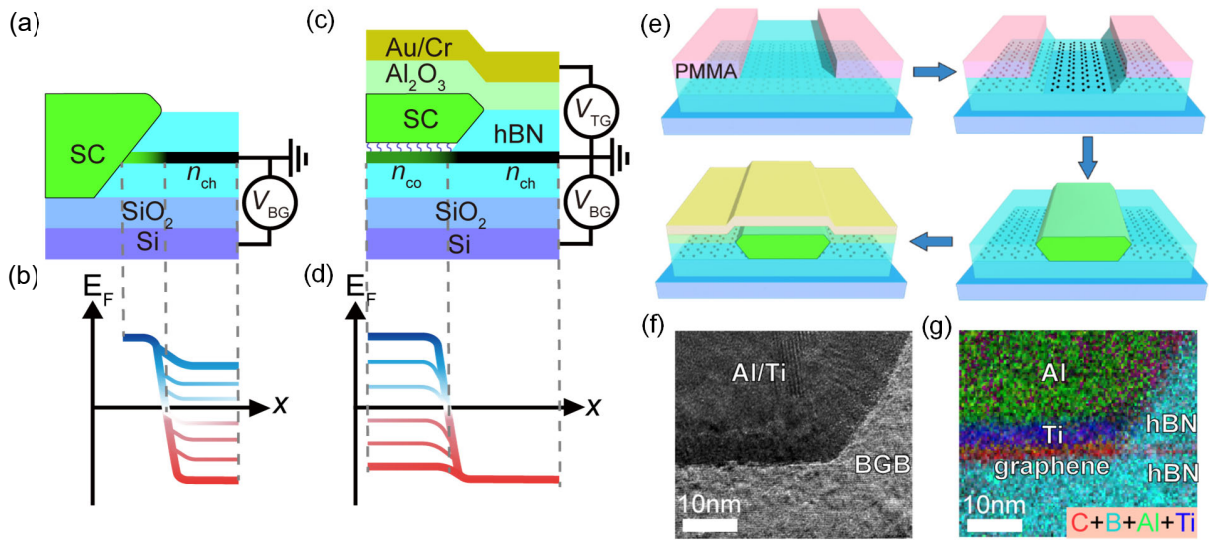

**Figure 1. Structure and fabrication procedures of two-dimensional(2D) superconducting (SC) contact.** (a)(b) Schematic illustrations of (a) 2D and (b) one-dimensional(1D) SC contact to graphene. (c)(d) Schematic illustration of the Fermi energy ($E_F$) in graphene near the (c) 2D and (d) 1D SC contact. (e) Fabrication procedures for a 2D SC contact on a hexagonal boron nitride(hBN)/graphene/hBN heterostructure. (f) High-resolution transmission electron microscope (HRTEM) image of 2D SC contact. (g) Electron energy loss spectroscopy (EELS) of 2D SC contact.

Figure 1e illustrates the fabrication procedures for the 2D superconducting contact in this work. The key feature of our fabrication is the formation of a 2D contact on the graphene surface as shown in Figure 1c, which is distinct from the 1D contact scheme[12] shown in Figure 1a. We fabricated superconducting contacts to graphene encapsulated with hexagonal boron nitride (hBN), which is stacked on a highly doped silicon (Si) wafer covered with a 300 nm of silicon dioxide ($SiO_2$) dielectric layer. The device shapes and electrodes are defined using conventional electron beam lithography. Immediately after selectively etching the top hBN of contact area without etching graphene, superconducting materials were deposited by either electron beam evaporation (for Ti, Al) or DC sputtering (for NbN, Nb, and MoRe) using the same etching mask. We selectively etched out the top hBN using a mild plasma of $CF_4$ with power of 10 W at 250 mTorr (Covance, Femtoscience). $CF_4$ plasma at relatively high pressure leads to isotropic chemical reaction which provides high etching selectivity[19], as opposed to anisotropic physical bombardment. The etching rate for hBN under this condition is 2~4 nm/s while monolayer graphene remains unetched even after an hour of etching, which manifests virtually infinite etching selectivity of hBN over graphene. After making contacts, an about 32 nm thick aluminum oxide layer is deposited through the atomic layer deposition (ALD), followed by the deposition of the top-gate electrode. The high-resolution transmission electron microscope (HRTEM) image in Figure 1f and the electron energy loss spectroscopy (EELS) image in Figure 1g show that the graphene layer remains in-between deposited superconducting material and the bottom hBN layer. In this 2D contact structure, the charge concentration in the contact area of graphene beneath the superconductor ($n_{co}$) can be controlled by the backgate voltage ($V_{BG}$) as $n_{co} = \alpha V_{BG}$, while that in the channel area of graphene ($n_{ch}$) can be controlled by the combination of the topgate voltage ($V_{TG}$) and $V_{BG}$ as $n_{ch} = \beta V_{BG} + \gamma V_{TG}$, where $\alpha$ and $\beta$ represent the gate coefficient for backgate and $\gamma$ represents the gate coefficient for topgate, respectively. Thus, $E_F$ of both regions can be independently controlled, which is in sharp contrast to the 1D contact case where $E_F$ near the contact is always fixed to a positive value (Figure 1b).

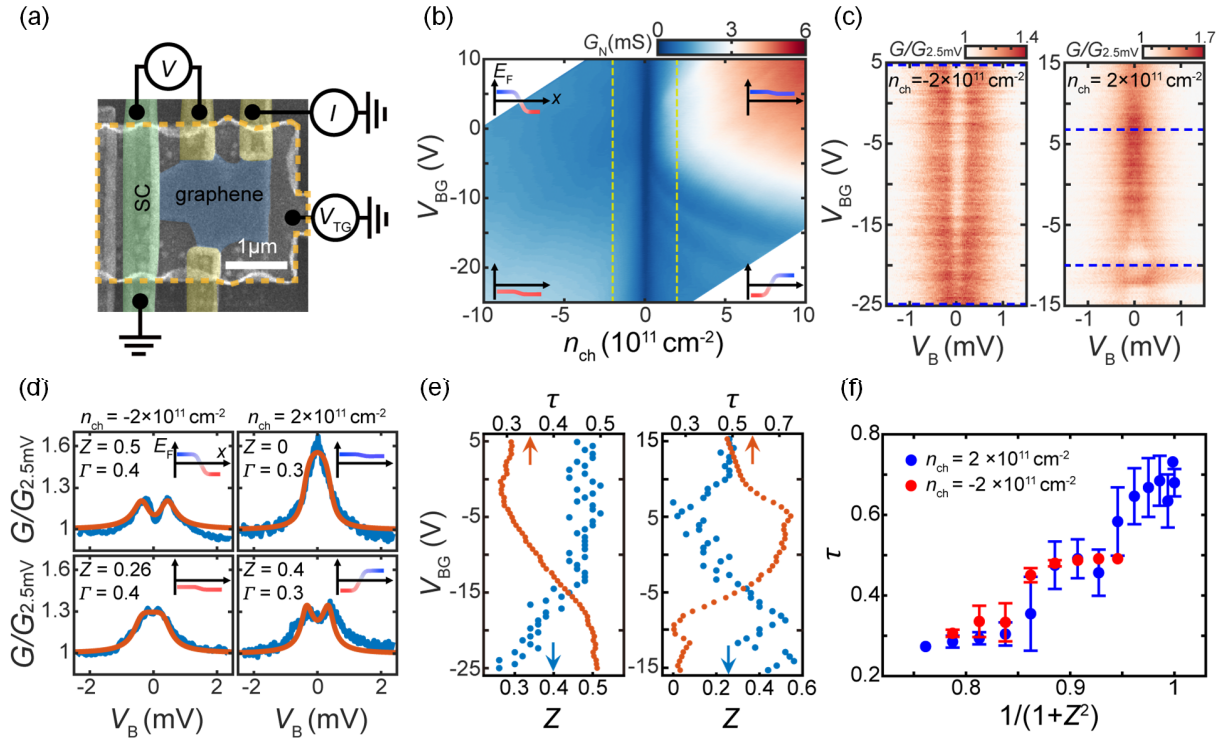

**Figure 2. 3-probe measurement of two-dimensional superconducting contact.** (a) False-colored scanning electron microscope image of the device with the measurement configuration. A NbN/Nb/Ti superconducting electrode (green) and Au/Cr normal electrodes (yellow) contact the graphene (blue). The orange dashed line indicates the topgate electrode. (b) Normal state conductance ($G_N$) as a function of the charge concentration in the channel area of graphene ($n_{ch}$) and applied backgate voltage ($V_{BG}$) at the temperature of $T = 10.8$ K. Two vertical yellow dashed lines correspond to $n_{ch} = -2\times10^{11}$ cm$^{-2}$ (left) and $n_{ch} = 2\times10^{11}$ cm$^{-2}$ (right), respectively, for (c). Insets in the four corners: schematic illustrations of Fermi energy ($E_F$) of graphene near the contact in the coordinate $x$ along the device length direction corresponding to the four polarity configurations. (c) Differential conductance normalized by the data taken at the bias voltage of $V_B = 2.5$ mV ($G_{2.5mV}$) at $n_{ch} = -2\times10^{11}$ cm$^{-2}$ (left) and $n_{ch} = 2\times10^{11}$ cm$^{-2}$ (right) as a function of $V_B$ and $V_{BG}$ at $T = 20$ mK. Blue dashed lines correspond to each panel of (d). (d) Line cuts of (c) at $V_{BG} = 6.8$ V (upper right), -10 V (lower right) at $n_{ch} = 2\times10^{11}$ cm$^{-2}$ and $V_{BG} = 4.925$ V (upper left), -25 V (lower left) at $n_{ch} = -$

$2\times10^{11}$ cm$^{-2}$, respectively. The red curves correspond to the Blonder-Tinkham-Klapwijk (BTK) fitting with the barrier strength $Z$, energy broadening $\Gamma$ and superconducting gap $\Delta$ = 0.45 meV. (e) Contact transmission probability $\tau$ and barrier strength $Z$ as a function of $V_{BG}$. (f) Correlation between $\tau$ and $1/(1+Z^2)$.

We characterized 2D superconducting contacts by measuring normal state conductance ($G_N$) and normalized differential conductance ($G/G_{2.5mV}$), which directly represent the transmission efficiency and Andreev reflection efficiency, respectively. We measured in a 3-probe configuration as depicted in Figure 2a. Provided that the electrons in encapsulated graphene undergo ballistic transport, 3-probe resistance would be dominated by the superconducting contact resistance. Normal state conductance ($G_N$) is measured at $T$=10.8K which is slightly lower than the critical temperature of NbN electrode where the contributions from conductance enhancement due to Andreev reflection and from the electrode resistance vanish (Figure. 2b). The backgate-tunability of $n_{co}$ is readily confirmed by the fact that $G_N$ of the 1$^{st}$ and the 3$^{rd}$ quadrants ($n_{co} \times n_{ch} > 0$) is higher than that of the 2$^{nd}$ and the 4$^{th}$ ones ($n_{co} \times n_{ch} < 0$). A polarity-reverting potential barrier exists in the 2$^{nd}$ and the 4$^{th}$ quadrants but not in the 1$^{st}$ and the 3$^{rd}$ quadrant as shown in the insets of Figure 2b. This backgate-tunability also enables in-situ control of the barrier strength $Z$ of normal-metal/superconductor interface in the Blonder-Tinkham-Klapwijk (BTK) model[1]. Figure 2c shows the bias-voltage dependence of normalized differential conductance $G/G_{2.5mV}$ with varying $V_{BG}$ at different polarities of $n_{ch}$. For $V_{BG}$ when the contact area and the channel area of graphene have different polarities, $G/G_{2.5mV}$ curve shows a dip near the zero-bias voltage ($V_B = 0$ V), indicating a high barrier strength. As $V_{BG}$ is adjusted such that the contact area and the channel area of graphene share the same polarity, the dip near zero-bias gradually transforms into a peak, signifying low barrier strength. This is more clearly revealed in Figure 2d, where each $G/G_{2.5mV}$ curve at four different graphene polarity configurations is quantitatively analyzed with the BTK model.

The most notable difference compared to the 1D SC contact case (Figure S1) is that, for a p-doped graphene channel ($n_{ch} = -2\times10^{11}$ cm$^{-2}$), the dip near $V_B = 0$ V gradually changes into a peak when a sufficiently large negative backgate voltage ($V_{BG} = -25$ V) is applied. In addition, for an n-doped graphene channel ($n_{ch} = 2\times10^{11}$ cm$^{-2}$), the barrier strength $Z$ drops to zero, and $G/G_{2.5mV}$ at $V_{BG} = 0$ V exceeds 1.6 at an optimal $V_{BG}$ of 6.8 V. This suggests that the potential barrier between the contact area and the channel area of graphene vanishes as $n_{co}$ is closely tuned to match $n_{ch}$. Figure 2e shows the $V_{BG}$ dependence of $Z$ extracted from the BKT model fitting and contact transmission probability $\tau$ calculated from $G_N$ as $\tau = G_N/G_Q$ with $G_Q = (4e^2/h)\times W/(\pi/n_{ch})^{1/2}$ and the width of the contact $W$. $Z$ and $\tau$ are anti-correlated for the whole range of measured $V_{BG}$. In the BTK model[20], $G_N$ and $Z$ have an relationship of $G_N = Ne^2/h \times 1/(1+Z^2)$ with the number of modes $N$, and this relationship can be confirmed by plotting $\tau$ and $1/(1+Z^2)$ as shown in Figure 2f.

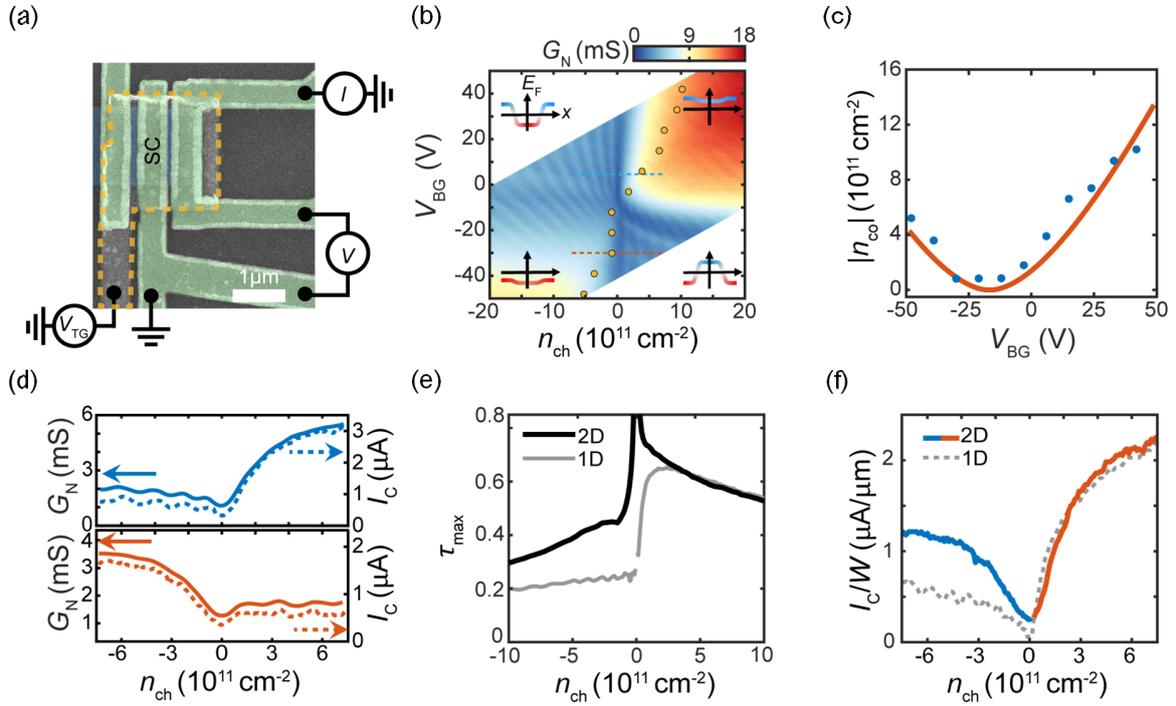

**Figure 3. Transport characteristics of a graphene Josephson junction (GJJ) with two-dimensional (2D) superconducting contact.** (a) False-colored scanning electron microscope image of the GJJ device fabricated with 2D superconducting molybdenum/rhenium (MoRe) contacts with measurement configuration. (b) Normal state conductance ($G_N$) as a function of charge concentration in the channel area of graphene ($n_{ch}$) and applied backgate voltage ($V_{BG}$) at temperature of $T$ = 8.5 K. The yellow dots correspond to $n_{ch}$ where the contact transmission probability becomes maximum ($\tau_{max}$) for given $V_{BG}$. Insets in the four corners: schematic illustrations of the Fermi energy ($E_F$) in the coordinate $x$ along the device length direction corresponding to the four polarity configurations. (c) Estimated $n_{co}$ as a function of $V_{BG}$ with a fit described in the main text. (d) $G_N$ and the critical current ($I_C$) of the GJJ as a function of $n_{ch}$ for $V_{BG}$ = 5 V (upper panel) and $V_{BG}$ = -30 V (lower panel) (e) Comparison of $\tau_{max}$ for 2D contact to 1D cases as a function of $n_{ch}$. (f) Comparison of width-normalized $I_C$ as a function of $n_{ch}$ for 1D and 2D contact cases. 2D contact data is taken at $V_{BG}$ = -30 V(blue) and $V_{BG}$ = 5 V(red).

We study how the backgate-tunability of the contact area of graphene improves the quality of graphene Josephson junction (GJJ) (Figure 3). As shown in Figure 3b, $G_N$ of GJJ follows a similar trend with that of 3-probe measurement shown in Figure 2b as it is higher in the 1st and 3rd quadrants than that in the 2nd and 4th quadrants. Additionally, it shows oscillatory behavior due to the Fabry-Perot (FP) interference more evidently in the 2nd and 4th quadrants than that in the 1st and 3rd quadrants. This contrasts with the GJJ fabricated with 1D contact, which shows FP interference in the 2nd and 3rd quadrants where $n_{co}$ is always negative (Figure S2). This quadrant dependent oscillatory behavior also supports that $n_{co}$ is controlled by $V_{BG}$ as it results from the corresponding $E_F$ configurations depicted in the insets of Figure 3b. Note that the

FP cavity is formed by p-n junctions between the channel area of graphene and the contact area of graphene, unlike the other reported cases[21] where the cavity is formed by p-n junctions within the channel area of graphene. Assuming that $\tau$ is maximized when $n_{ch}$ equals $n_{co}$, we extract $n_{co}$ as a function of $V_{BG}$ (orange circles in Figure 3b). In Figure 3c, we constructed a fitting for the $V_{BG}$ dependence of $n_{co}$, taking into account electrostatic gating and screening effect[18] with a work function mismatch of 50 meV and a distance between electrode and graphene of 0.3 nm. Similar to the case of conductance enhancement at the superconducting contact discussed in Figure 2, the critical current ($I_C$) of the GJJ also improves with $G_N$. The correlation between $G_N$ and $I_C$ is illustrated in Figure 3d, where both values are higher when $n_{co}$ and $n_{ch}$ have the same polarity. The comparison of maximum contact transmission probability $\tau_{max}$ (Figure 3e) and width-normalized $I_C$ (Figure 3f) with the case of 1D contact directly demonstrates the improvement of GJJ quality by tuning $n_{co}$ with 2D contact scheme. Both $\tau_{max}$ and $I_C/W$ exhibits up to twofold improvement for $n_{ch} < 0$.

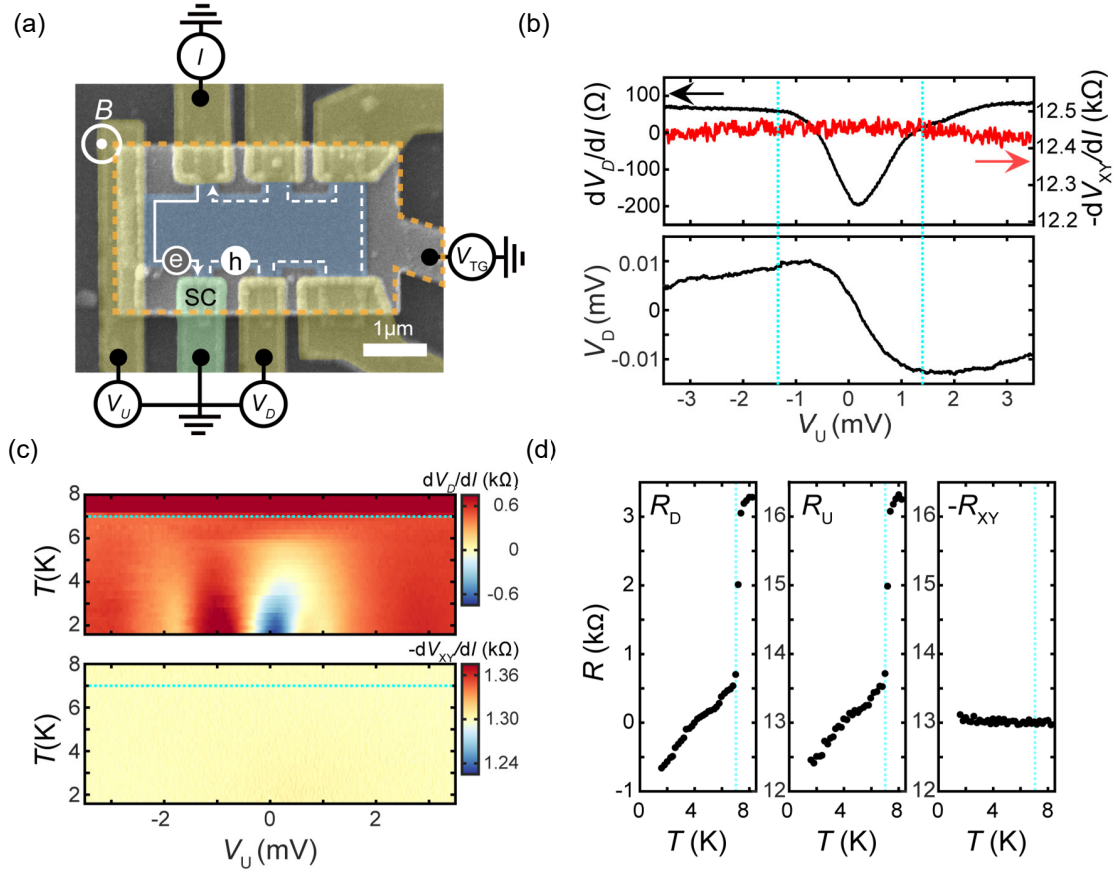

**Figure 4. Crossed Andreev conversion in the quantum Hall state of negative filling factor $\nu$ = -2.** (a) False-colored scanning electron microscope image of graphene Hall bar device fabricated with two-dimensional (2D) superconducting molybdenum/rhenium (MoRe) contacts with measurement configuration. (b) Upper panel: differential downstream resistance ($dV_D/dI$) (black) and differential Hall resistance ($-dV_{XY}/dI$)

(red) as a function of upstream bias voltage ($V_U$) at magnetic field $B$ = 4.559 T, backgate voltage $V_{BG}$ = -30 V and topgate voltage $V_{TG}$ = 6.298 V. Lower panel: downstream voltage ($V_D$) as a function of $V_U$. Sky-blue vertical dashed lines indicate the superconducting gap of MoRe ($\Delta_{MoRe}$). (c) Color-coded maps of d$V_D$/d$I$ (upper panel) and -d$V_{XY}$/d$I$ (lower panel) as a function of $V_U$ and temperature $T$ at $B$ = 3 T, $V_{BG}$ = -10 V and $V_{TG}$ = 0.9747 V. Sky-blue horizontal dashed lines indicate the critical temperature of MoRe electrodes. (d) Downstream resistance ($R_D = V_D/I$), upstream resistance ($R_U = V_U/I$) and Hall resistance ($-R_{XY} = (V_U-V_D)/I$) as a function of $T$ corresponding to the zero-bias ($V_U$=0) cut of (c). Sky-blue vertical dashed lines indicate the critical temperature of MoRe electrodes.

To showcase the effectiveness of the 2D contact scheme in inducing superconducting correlations within the quantum Hall edge state, we present the experimental measurements on the crossed Andreev conversion process in quantum Hall edge state for negative filling factor of $v$ = -2. When the quantum Hall edge state encounters a superconducting electrode, an upstream electron can undergo Andreev reflection, converting into a downstream hole (Figure 4a), which manifests as a negative downstream resistance ($R_D$). Therefore, $R_D$ not only represents the contact resistance between quantum Hall edge state and the drain electrode, but also serves as an indicator of the effectiveness of the hybridization between superconducting electrode and the quantum Hall edge state by giving a negative value[6,7,8,9,22,23]. However, the Andreev process at negative filling factor has not been reported due to low contact transparency caused by the formation of $v$ = 0 near superconducting contact. Here, we fabricated a graphene Hall bar device with MoRe superconducting 2D contacts (Figure 4a) which shows $R_D$ (Figure S3) to be less than 100 $\Omega$ up to a magnetic field of 10 T for the filling factors $v$ = -2 and $v$ = 2, demonstrating high transparency in both polarities. In the upper panels of Figure 4b, downstream differential resistance (d$V_D$/d$I$) shows a negative value when the upstream bias voltage ($V_U$) is within the superconducting gap of MoRe electrode ($\Delta_{MoRe}$ ~ 1.4 meV) for a negative filling factor of $v$ = -2. For complementary verification, we also measured downstream DC voltage (lower panel of Figure 4b) as a function of $V_U$, which clearly demonstrates the negative slope within $\Delta_{MoRe}$. The BCS-like temperature ($T$) dependence for the negative d$V_D$/d$I$ (upper panel of Figure 3c) indicates that the negative value indeed arises from superconductivity, while the differential Hall resistance -d$V_{XY}$/d$I$ (lower panel of Figure 3c) remains its quantized value of $h/2e^2$ regardless of the changes in $T$ and $V_U$. Both upstream ($R_U = V_U/I$) and downstream ($R_D = V_D/I$) resistance continuously decrease below the critical temperature of MoRe ($T_{MoRe}$ = 7 K) electrode (Figure 4d).

A question may arise regarding how the charge concentration of graphene that is fully contacted to metal can be controlled merely by applying gate voltage. In contrast to the 1D edge contact case where sp2 bonding of graphene is broken at the etched edge, the 2D surface of graphene lacks dangling bonds so the coupling between the superconductor and graphene would be weaker. This leads to weaker doping on the graphene such that the doping can be effectively controlled by electrostatic gating. Despite the weak bonding between the superconductor and graphene, the number of conduction channels corresponds to the number of π bonds in contact with the superconducting electrodes. These numerous conduction channels in the 2D contact result in sufficiently high conductance, leading to a robust superconducting proximity effect. The measured contact resistance between the superconductor and graphene (Figure S4) is negligibly small, indicating that the finite contact transmission probability $\tau$ we discussed above mainly originates from the interface between the channel area of graphene and contact area of graphene, not between the superconducting electrode and the contact area of graphene.

Previous studies have shown that fluorinating graphene using fluorine-based plasma significantly decreases the conductance of the graphene channel by several orders of magnitude, resulting in a loss of gate-tunability[24,25]. However, our hBN etching recipe does not cause a significant deterioration on the exposed surface of graphene, thereby preserving the gate-tunability (Figure S5). This is likely due to the relatively high pressure during our hBN etching process, which results in a shorter mean free path for the $CF_4$ plasma and less implantation of fluorine atoms on the graphene surface.

This 2D contact works not only for superconducting electrodes but also for other contact materials commonly used with graphene such as Ti, Ta, MoRe and Cr (Figure S6). Additionally, the automatic stop of hBN etching at graphene, due to its virtually infinite etching selectivity, offers better yield for the devices with graphite gates, which frequently fail due to electrical shorts between the electrodes and graphite gates. Our proposed fabrication process may pave the way for more versatile superconducting hybrid devices featuring bipolar graphene with high contact transparency.

Acknowledgements

This research was supported by National Research Foundation of Korea (NRF) funded by the Korean Government (2020M3H3A1100839, RS-2024-00393599, RS-2024-00442710, RS-2024-00444725), ITRC program (IITP-2022-RS-2022-00164799) funded by the Ministry of Science and ICT, Samsung Science and Technology Foundation (SSTF-BA2101-06, SSTF-BA2401-03) and Samsung Electronics Co., Ltd (IO201207-07801-01). S.J. was supported by Basic Science Program through the National Research Foundation of Korea (NRF) funded by the Ministry of Education(2022R1A6A3A01086903). K.W. and T.T. acknowledge support from the JSPS KAKENHI (Grant Numbers 21H05233 and 23H02052) and World Premier International Research Center Initiative (WPI), MEXT, Japan.

**Supplementary information**

**for**

**Engineering superconducting contacts transparent to a bipolar graphene**

Seong Jang[1], Geon-Hyoung Park[1], Sein Park[1], Hyeon-Woo Jeong[1], Kenji Watanabe[2], Takashi Taniguchi[3] and Gil-Ho Lee[1]

[1] Department of Physics, Pohang University of Science and Technology, Pohang, Republic of Korea

[2] Research Center for Functional Materials, National Institute for Materials Science, Tsukuba, Japan

[3] International Center for Materials Nanoarchitectonics, National Institute for Materials Science, Tsukuba, Japan

1. Device fabrication

The stacks of hexagonal boron nitride (hBN)/graphene/hBN are prepared using the dry-transfer method with Elvacite[1] which undergoes a glass transition at a temperature of approximately 180 K. The stacks are annealed in vacuum at 773 K for 2 hours to remove bubbles and residues trapped between hBN and graphene. Subsequently, the topography of stack surface is scanned by atomic force microscopy to locate regions without any bubble or wrinkle, which is then fabricated into Josephson junctions or Hall bar devices. We use PMMA A4 or ARP6200 for the electron beam resist with a thickness of 240 nm approximately. We adopt different etch recipes for hBN and graphene. The CF4 plasma process is conducted for hBN etching within a chamber pressure range of 200~300 mTorr and with the plasma power of 10 W. The etching rate of hBN typically falls within the range of 2-4 nm/s. The etching time is adjusted to sufficiently etch the top hBN flake, considering its thickness. Alternatively, graphene is etched using $O_2$ plasma within the chamber pressure range of 200~300 mTorr and with a plasma power of 10 W, resulting in an etching rate of approximately one graphene layer every five seconds. We deposit superconducting materials using the same electron beam resist mask to avoid any contamination on the exposed graphene which could deteriorate the contact between graphene and superconductor. For niobium nitride/niobium/titanium (NbN/Nb/Ti) contact, we deposit 5 nm of titanium by electron beam evaporation, followed by the deposition of 5 nm of Nb and 60 nm of NbN by DC sputtering with a power of 300 W and a pressure of 3.4 mTorr, resulting in superconducting critical temperature of 11.5 K. MoRe is deposited by DC sputter at a power of 300 W and a pressure of 3.4 mTorr. We deposit 75 nm of MoRe for the graphene Josephson junction device shown in Figure 3 and 60 nm of MoRe for the Hall bar device shown in Figure 4, resulting in the superconducting critical temperature of 8.46 K and 7 K, respectively. For the control experiment, we fabricate graphene Josephson junctions with one-dimensional (1D) edge contact, where we etched out not only the top hBN, but also the graphene and bottom hBN. Except for the etching process, all fabrication steps are the same as used in the 2D contact fabrication process.

2. measurement

The normal state conductance ($G_N$) is measured using a lock-in amplifier with an excitation current of 10 nA or 100 nA and a frequency of 13.777 Hz at a temperature slightly lower than the critical temperature of superconducting electrodes, where conductance enhancement from Andreev reflection is negligible. The critical current ($I_C$) of the graphene Josephson junction (GJJ) is measured at the base temperature (~20 mK) of cryogen-free dilution refrigerator (Bluefors LD400) equipped with Thermocoax cables and multi-stage of pi and low pass filters. The DC bias current is applied to the GJJ by a Yokogawa DC voltage source and a load resistor. The voltage across the GJJ is then amplified with a gain of 1000 by an Ithaco voltage preamplifier and subsequently measured by a Keithley 2000 multimeter. For the measurement in the quantum Hall regime, a mixed DC and AC voltage is applied to either a 1 or 10 MΩ resistor to supply current to the device, and the voltage drop is measured the voltage drop with a lock-in amplifier.

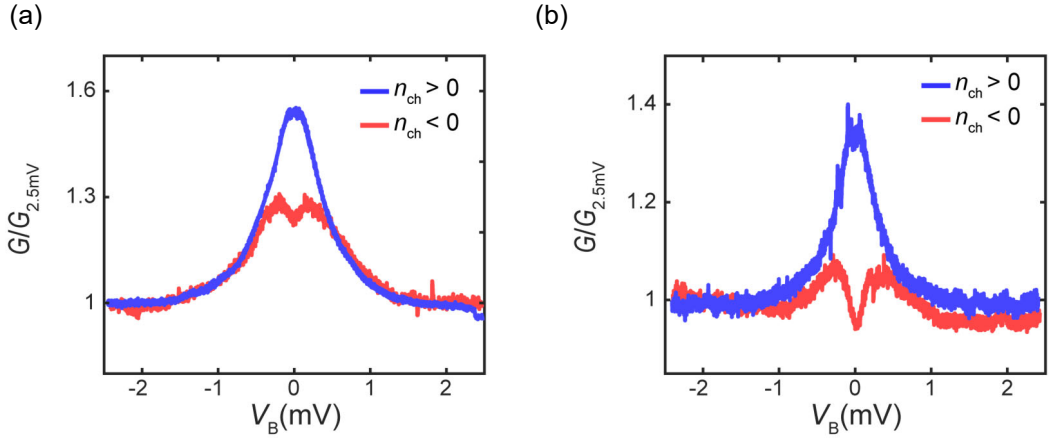

**Figure S1** Normalized differential conductance of (a) two-dimensional (2D) superconducting contact and (b) one-dimensional (1D) superconducting contact as a function of bias voltage for both polarities of graphene channel charge concentration $n_{ch} > 0$ and $n_{ch} < 0$. The data were taken at $V_{BG}$ = 0 V, $V_{TG}$ = 0 V ($n_{ch} > 0$) and $V_{BG}$ = -25 V, $V_{TG}$ = 3.59 V ($n_{ch} < 0$) for the 2D contact, $V_{BG}$ = 10 V, $V_{TG}$ = 0 V ($n_{ch} > 0$) and $V_{BG}$ = -20 V, $V_{TG}$ = 0 V ($n_{ch} < 0$) for the 1D contact, respectively.

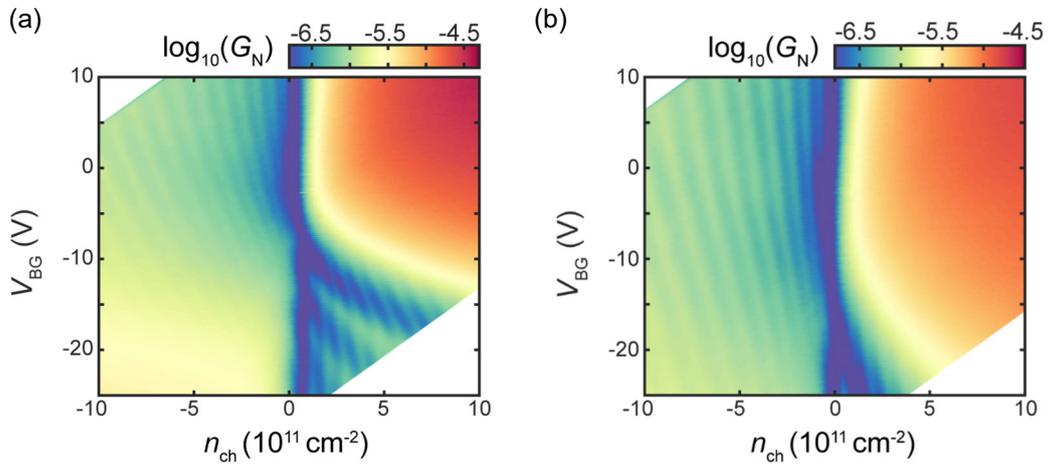

**Figure S2** Normal state conductance of graphene Josephson junctions fabricated with (a) two-dimensional superconducting contact and (b) one-dimensional superconducting contact as a function of the charge concentration in the channel area of graphene and backgate voltage measure at a temperature of 10.8 K which is slightly lower than the critical temperature of niobium nitride/niobium/titanium contact. The data are displayed on a logarithmic scale to enhance the visibility of Fabry-Perot interference.

## 3. Comparison to the one-dimensional (1D) superconducting contact devices

We fabricated and measured graphene Hall bar devices and Josephson junctions with 1D superconducting contacts to compare them with the 2D superconducting contact devices discussed in the main text. While the normalized differential conductance ($G/G_{2.5mV}$) of 2D contact Hall bar device (Figure S1 (a)) shows a peak near the zero bias voltage ($V_B$ = 0 V) regardless of the charge concentration of the channel area of graphene ($n_{ch}$), the 1D contact device exhibits a dip (Figure S1 (b)) when $n_{ch}$ is negative. This result indicates that the charge concentration in the contact area of graphene ($n_{co}$) is not tunable in devices with 1D contacts, unlike those with 2D contact.

This behavior is also evident in the Fabry-Perot (FP) interference in graphene Josephson junctions (GJJ). Figure S2 compares the normal state conductance ($G_N$) of graphene Josephson junctions with 2D contact (Figure S2 (a)) and 1D contact (Figure S2 (b)). Unlike the 2D contact GJJ which shows FP interference in the 2nd and 4th quadrants, 1D contact GJJ exhibits FP interference in the 2nd and the 3rd quadrant. This further supports the notion that $n_{co}$ is not tunable in the device with 1D contact.

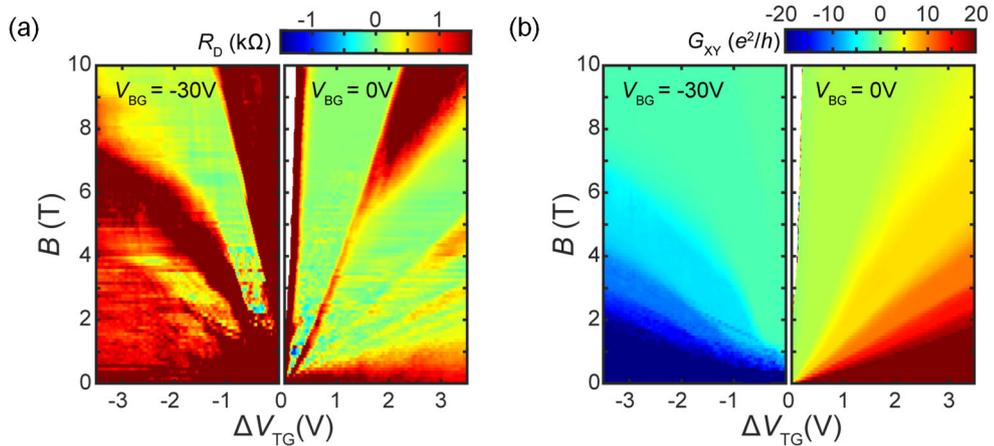

**Figure S3** (a) Fan diagram of downstream resistance ($R_D$) when the contact area of graphene is positively doped (left) and negatively doped (right). (b) Fan diagram of Hall conductance when the contact area of graphene is positively doped (left) and negatively doped (right).

## 4. Fan measurement of Hall bar device with 2D superconducting contact

We measured the Fan diagram of the graphene Hall bar device, as shown in Figure 4 of the main text, to confirm that the contact between the quantum Hall state and the superconducting electrode is transparent enough to allow for the Andreev process. For negative filling factors, we applied $V_{BG}$ = -30 V to tune $n_{co}$ to a negative value. For positive filling factors, no backgate voltage was applied. The downstream resistance ($R_D$), which serves as the contact resistance between quantum Hall state and the drain electrode, was measured to be less than 100 Ω for both filling factors of $v$ = -2 and $v$ = 2 (Figure S3 (a)), up to the magnetic field of 10 T while the quantum Hall conductance ($G_{XY}$) remained constant (Figure S3 (b)).

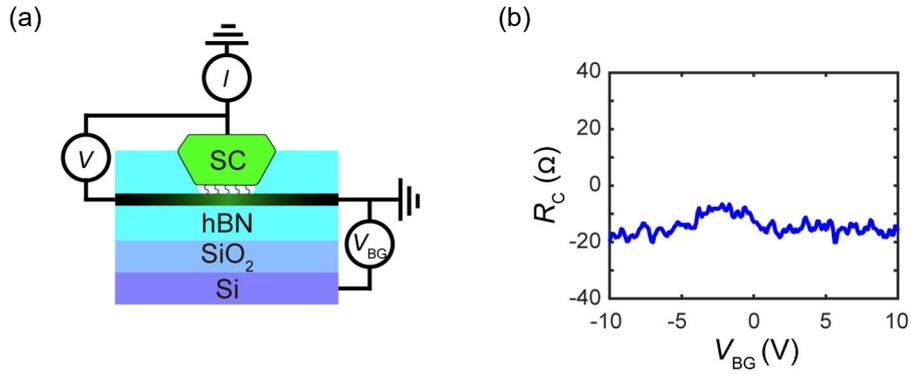

**Figure S4** (a) Schematics of measurement configurations for the contact resistance between the superconductor and the two-dimensional graphene surface. (b) Measured contact resistance ($R_C$) as a function of backgate voltage ($V_{BG}$) at a temperature of 300K.

5. Contact resistance between superconducting electrode and graphene

The measured resistance in the devices shown in Figure 2 in the main text comprises both the resistance between the channel area of graphene and the contact area of graphene, and the resistance between the superconducting electrode and the contact area of graphene. To clarify which component contributes more significantly, we measured the latter using the configuration depicted in Figure S4 (a), where the measured voltage difference arises solely from the contact area of graphene and the superconducting electrode. The measured resistance even goes below zero (Figure S4 (b)). This observation supports the conclusion that the resistance of the devices discussed in the main text predominantly originates from the resistance between the channel area of graphene and the contact area of graphene.

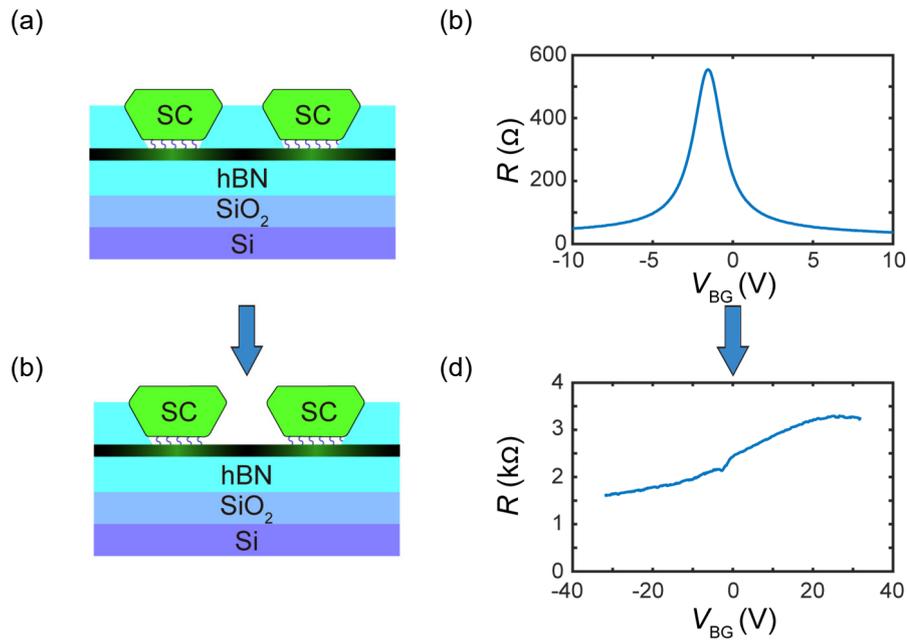

**Figure S5** (a) Measured resistance as a function of backgate voltage ($V_{BG}$) of graphene device encapsulated by hexagonal boron nitride (hBN) at a temperature of 300 K. (b) Measured resistance as a function of backgate voltage ($V_{BG}$) of graphene device after etching the top hBN layer.

6. Effect of fluorination

A number of studies [2,3,4,5] have demonstrated that the fluorine-based processes can fluorinate the graphene surface. This fluorination significantly reduces the conductivity of graphene by several orders of magnitude and causes the loss of gate-tunability[4,5]. Our hBN etching recipe also reduces the conductivity of graphene (Figure S5 (a)(b)), however, the gate-tunability is still preserved (Figure S5 (b)). Additionally, there is Dirac point shift indicating that our hBN etching recipe positively dopes the graphene. Such positive doping may neutralize the strong negative doping from contact material, resulting the Dirac point of contact area of graphene being closer to the neutral point, allowing us to control its polarity.

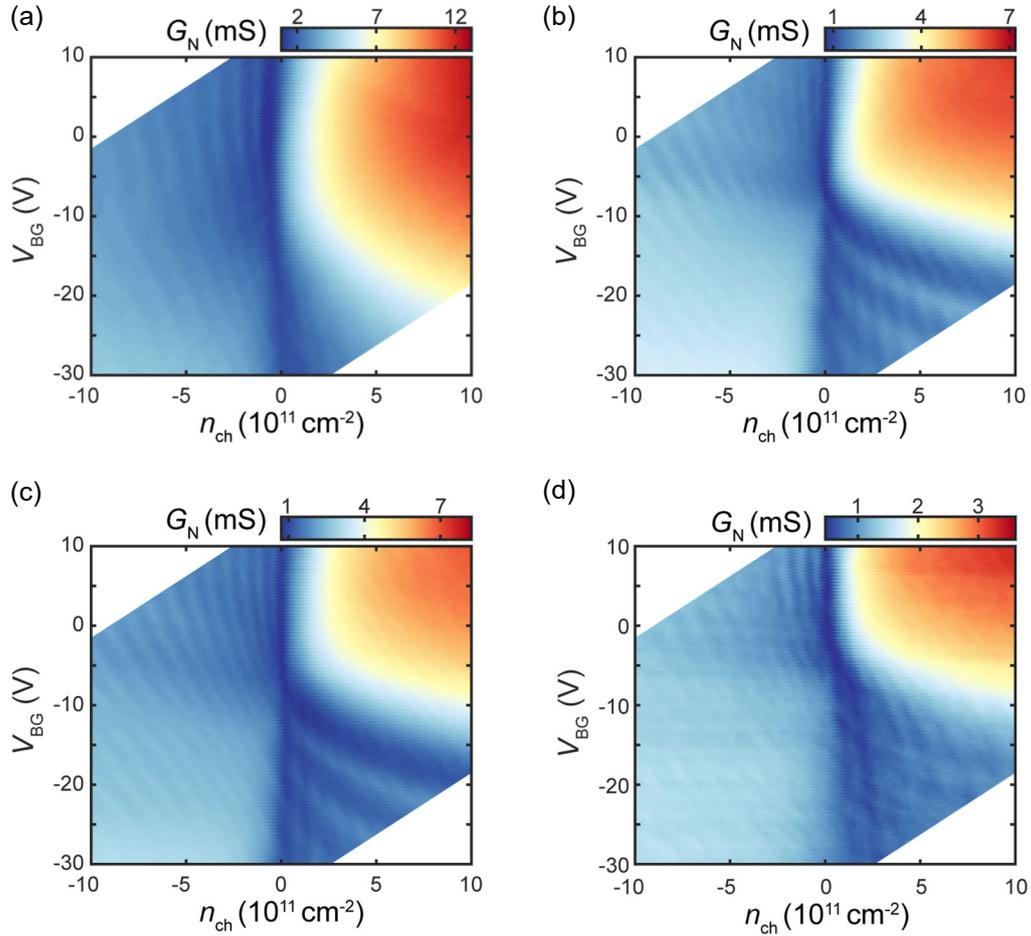

**Figure S6** Normal state conductance ($G_N$) as a function of the charge concentration in the channel area of graphene ($n_{ch}$) and applied backgate voltage ($V_{BG}$) for various contact materials : (a) Al/Ti (b) Ta (c) TaN/Ta (d) Au/Cr.

7. Applicability of 2D contact to various materials

We fabricated and measured graphene Josephson junctions with NbN/Nb/Ti, MoRe (main text), Al/Ti (Figure S6 (a)), Ta (Figure S6 (b)), TaN/Ta (Figure S6 (c)) and Au/Cr (Figure S6 (d)). All the tested materials exhibited similar behavior, supporting the conclusion that this 2D contact fabrication procedure can be universally applied.